\documentstyle[prd,graphicx,aps]{revtex}
\begin{document}

\title{Brane Variation Dirac Style}
\author{David Karasik and Aharon Davidson}
\address{Physics Department, Ben-Gurion University of the
Negev, Beer-Sheva 84105, Israel\\
\textsf{(karasik@bgumail.bgu.ac.il, davidson@bgumail.bgu.ac.il)}}
\maketitle

\begin{abstract}
Dirac's method for variations of a brane embedded in co-dimension
one is demonstrated. The variation in the location of the brane
invokes a rest frame formulation of the 'sandwiched' brane action.
We first demonstrate the necessity of this method by re-deriving
Snell's law. Second, we apply the method to a general
$N$-dimensional brane
    embedded in co-dimension one bulk in the presence of gravity.
We re-derive the brane equations:
    (i) Israel junction condition,
    (ii) Energy/momentum conservation on the brane, and
    (iii) Geodetic-type equation for the brane.
\end{abstract}

\section{Introduction}
In this paper we demonstrate Dirac's variation method
\cite{dirac} on a brane. Dirac's motivation was to describe the
electron as a bubble living in a background
of electromagnetic field. Since the bubble has one
dimension less then the background manifold,
it is actually a boundary between the inner and outer
parts of the surrounding manifold. A self consistent
model is obtainable by the principle of least Action,
the action must not change under small variations
both in the electromagnetic field and in the location of the bubble.

Dirac claims that the naive way for varying the location of the
bubble is wrong. This naive way is : parameterizing the bubble by
the coordinates $x^{\mu}$ and the surrounding manifold by the
coordinates $y^{A}$, the variation in the location of the bubble
is naively $\delta y^{A}(x)$, but this will lead to the wrong
equations. The right way to describe this variation is more
complicated, this calls for a new coordinate system $z^{a}$ in the
surrounding manifold. In the $z$-system the bubble's location is
fixed, while the variation in the location is done by varying the
whole $y$-system with respect to the $z$-system. Therefore the
action must be written in terms of the $z$-system, and the
canonical fields are the electromagnetic field and the
$y$-coordinate system.

The purpose of this paper is to demonstrate Dirac's method, and to
generalize it to include gravity. This paper is built as follows,
in section \ref{sec:snell} we give a simple example for this
method. This example is Snell's law of geometrical optics. We show
that the naive variation leads to the wrong condition, and only
Dirac's method gives the correct law. In section \ref{sec:general}
we start from the general action in $(N+1)$ dimensions, which
includes gravity in $(N+1)$-dimensions and an $N$-dimensional
brane. We perform the variation with respect to the $z$-system
where the brane is fixed, and derive the equations of motion. The
equations of the brane are not new. These are; (i) Israel junction
conditions \cite{israel}, (ii) generalized energy-momentum
conservation on the brane, and, (iii) the geodetic equation of the
brane \cite{carter}.

\section{A Simple Example : Snell's Law}
\label{sec:snell} 
\begin{figure}[h]
    \begin{center}
    \includegraphics[scale=0.5]{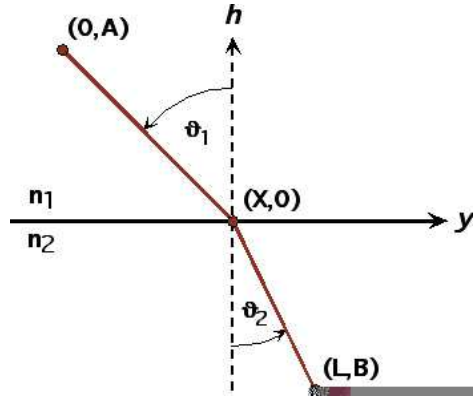}
    \end{center}
    \caption{A light ray propagating from $(0,A)$ to $(L,B)$,
    crossing the point $(X,0)$.}
    \label{snellfig}
\end{figure}

In geometrical optics we describe the light as
rays propagating with the speed $v=\frac{c}{n}$, where $n$ is the
index of refraction. We start with two media with coefficients
$n_{1}$,$n_{2}$, a light source is placed in the first medium
while the observer is in the second, see Fig. \ref{snellfig}.
The point $X$ where the ray crosses the surface between the
two media is the analog of the brane
(here it has a zero dimension).
The surrounding manifold is the $y$-axis, the brane
separates the axis into two parts.
The canonical variable here is the height $h(y)$,
while the action is the total time, it can be written as
an integral over $y$
\begin{equation}
    I= \int_{0}^{X}dy\,\frac{n_{1}}{c}\sqrt{1+ (dh/dy)^{2}} +
     \int_{X}^{L}dy\,\frac{n_{2}}{c}\sqrt{1+ (dh/dy)^{2}}
    \label{snellaction}
\end{equation}
The naive variations are $\delta h(y)$ and $\delta X$.
At the end points and on the brane, the variation of
$h$ vanishes and therefore the variation of the action is
\begin{eqnarray}
    \delta I &=& - \frac{n_{1}}{c} \int_{0}^{X}dy\,\delta h
\frac{d}{dy}\left(\frac{dh/dy}{\sqrt{1+ (dh/dy)^{2}}}\right)
    + \frac{n_{1}}{c}\left.\delta X \sqrt{1+ (dh/dy)^{2}}
\right|_{X^{-}} \nonumber \\
     & - &  \frac{n_{2}}{c} \int_{X}^{L}dy\,\delta h\frac{d}{dy}
    \left(\frac{dh/dy}{\sqrt{1+ (dh/dy)^{2}}}\right)
 - \frac{n_{2}}{c}\left.\delta X \sqrt{1+ (dh/dy)^{2}}
\right|_{X^{+}}
    \label{wrongvari}
\end{eqnarray}
The equation of motion within each medium is simply
$\frac{dh}{dy}=const.$, which
means that the light propagates in straight lines.
The variation with respect to $X$
will lead to the equation
\begin{equation}
    n_{1}\left. \sqrt{1+ (dh/dy)^{2}} \right|_{X^{-}}
    = n_{2}\left. \sqrt{1+ (dh/dy)^{2}} \right|_{X^{+}}
\end{equation}
The geometrical relations
\begin{eqnarray}
    \cot\theta &=& \frac{dh}{dy} \\
    \frac{1}{\sin\theta} &=& \sqrt{1+(dh/dy)^{2}}
    \label{sintheta}
\end{eqnarray}
will lead us to the \textbf{wrong} relation
\begin{equation}
    \frac{n_{1}}{\sin\theta_{1}}= \frac{n_{2}}{\sin\theta_{2}}
    \label{wrongsnell}
\end{equation}
The reason for this wrong relation is that a variation
in $X$ will change the path of the ray everywhere and
therefore can not be regarded as an independent variation
\footnote{There is however an option to get Snell's law without
Dirac's method. That is to solve the equations for $h(y)$ with
the specific boundary conditions, to calculate the total action
as a function of $X$ and to impose $\frac{dI}{dX}=0$.}.

Now we will use Dirac's method for the variation.
Take another coordinate for the axis, and call it $z$.
In the $z$-axis, the location of the point does not change.
The action (\ref{snellaction}) is written as an integral over $z$
\begin{equation}
    I= \int_{0}^{\tilde{X}}dz\,\frac{n_{1}}{c}
\sqrt{(y')^{2}+(h')^{2}} +
     \int_{\tilde{X}}^{L}dz\,\frac{n_{2}}{c}\sqrt{(y')^{2}+ (h')^{2}}
    \label{snellaction2}
\end{equation}
The prime denotes differentiation with respect to $z$.
The canonical variables are $y(z)$ and $h(z)$,
while the limits of integration are fixed.
The variation is
\begin{eqnarray}
    \delta I &=& - \frac{n_{1}}{c} \int_{0}^{\tilde{X}}dz\,
    \left[\delta h\frac{d}{dz}
    \left(\frac{h'}{\sqrt{(y')^{2}+ (h')^{2}}}\right)
    +  \delta y\frac{d}{dz}
    \left(\frac{y'}{\sqrt{(y')^{2}+ (h')^{2}}}\right) \right]
    \nonumber \\
    & - &   \frac{n_{2}}{c} \int_{\tilde{X}}^{L}dz\,
    \left[\delta h\frac{d}{dz}
    \left(\frac{h'}{\sqrt{(y')^{2}+ (h')^{2}}}\right)
    +  \delta y\frac{d}{dz}
    \left(\frac{y'}{\sqrt{(y')^{2}+ (h')^{2}}}\right)\right]
     \nonumber \\
    & + & \frac{n_{1}}{c}\left.\delta y \frac{y'}{\sqrt{(y')^{2}+ (h')^{2}}}
    \right|_{\tilde{X}^{-}}
    - \frac{n_{2}}{c}\left.\delta y \frac{y'}{\sqrt{(y')^{2}+ (h')^{2}}}
    \right|_{\tilde{X}^{+}}
    \label{truevari}
\end{eqnarray}
The light propagates in straight lines in the $(y,h)$ plane,
since $\frac{dh}{dy}=\frac{h'}{y'}=const$,
while in the $z$-plane the motion might look very complicated.
The variation  $\delta y$ is continuous over the 'brane', that is
$\delta y^{-}=\delta y^{+}$. Using now the geometrical relations
(\ref{sintheta}), the true matching condition is therefore
\begin{equation}
    n_{1}\sin\theta_{1} = n_{2}\sin\theta_{2}
    \label{snell}
\end{equation}
This simple example demonstrates Dirac's method for
variations on the brane.

\section{The General Action}
\label{sec:general}
We would like to apply Dirac's method to a general
$N$-dimensional brane living in a background
with one extra dimension.
The general action for such a brane should look like
\begin{eqnarray}
    I &=& \int_{V_{1}}d^{N+1}y\,\sqrt{-G}
    \left[\frac{\cal R}{16\pi\bf G}+{\cal L}_{1}  \right]
        -\int_{B}d^{N}x\,\sqrt{-g}\frac{K_{1}}{8\pi\bf G}
    \nonumber \\
    &+& \int_{B}d^{N}x\,\sqrt{-g}\,L_{B}    \label{genaction}\\
    &+& \int_{V_{2}}d^{N+1}y\,\sqrt{-G}
    \left[\frac{\cal R}{16\pi\bf G}+{\cal L}_{2}  \right]
    -\int_{B}d^{N}x\,\sqrt{-g}\frac{K_{2}}{8\pi\bf G} \nonumber
\end{eqnarray}
\begin{itemize}
\item The embedding manifold has been separated
    into $V_{1,2}$ by the brane $B$.
\item The $(N+1)$-dimensional line element is
    $ds^{2}_{N+1}=G_{AB}dy^{A}dy^{B}$, while the line
element on the brane is
$ds^{2}_{N}=g_{\mu\nu}dx^{\mu}dx^{\nu}$.
\item ${\bf G}$ is the bulk gravitational coupling, i.e.
Newton's constant in $(N+1)$ dimensions.
\item The gravitational action in $V_{1,2}$ includes the
    $(N+1)$-dim integral of the scalar curvature ${\cal R}$ over
$V_{1,2}$, plus an $N$-dim integral over the brane
of the extrinsic curvature of the brane embedded in
$V_{1,2}$ that is $K_{1,2}$.
\item Matter Lagrangians in $V_{1,2}$ are ${\cal L}_{1,2}$.
\item The brane itself is characterized by the integral
    of the Lagrangian $L_{B}$. This Lagrangian might
    include the $N$-dim intrinsic curvature as well as
    $N$-dim matter fields. (The ordinary symbols are used for the
    brane objects while the script symbols are for the bulk objects.)
\item For simplicity we assume here that the extra dimension
    is space-like. If it is time-like there should be only
some sign changes, while if it is null a more
careful treatment is needed.
\end{itemize}
\subsection{Variation in the Rest Frame of the Brane }
According to Dirac's method, we must describe the action
in term of a special coordinate system, where the brane
is fixed. Let us denote such a system with the coordinates
$z^{a}$. (Upper case Latin indices are saved for the $y$
system, while lower case Latin indices are for the $z$
system. Greek indices are the brane indices).
The transformation of tensors from one system to another
is done in the usual way with the bi-tensorial object
$y^{A}_{\,,a}=\frac{\partial y^{A}}{\partial z^{a}}$,
for example the metric
\begin{equation}
    G_{ab}=y^{A}_{\,,a}y^{B}_{\,,b}G_{AB}~.
    \label{zmetric}
\end{equation}
In general one can adopt a different coordinate system for each
side of the brane. The following geometrical relations are
applicable for each side of the brane separately. The brane we are
working with is a hypersurface in the surrounding manifold. The
location of the brane is defined by the function $f(z)=0$. The
normal outward-pointing unit vector is
\begin{equation}
    n_{a}\propto f_{,a}\; ; \; G^{ab}n_{a}n_{b}=1
    \label{normal}
\end{equation}
The first fundamental form is
$h^{a}_{b}=\delta^{a}_{b}-n^{a}n_{b}$, it is a projection
operator onto the tangent space of the brane, such that
\begin{equation}
    h^{a}_{b}n_{a}=0\; ; \; h^{a}_{c}h^{c}_{b}=h^{a}_{b}
    \label{fff}
\end{equation}
The second fundamental form $K_{ab}$ is the extrinsic curvature
of the brane embedded in the external manifold.
\begin{equation}
    K_{ab}=-\frac{1}{2}h^{c}_{a}h^{d}_{b}
(n_{c;d}+n_{d;c})~.
\label{excurv}
\end{equation}
Two important relations relate the extrinsic curvature of the
brane, the intrinsic curvature of the brane and the curvature
of the surrounding manifold. These are the Gauss relation
\begin{equation}
    R= {\cal R} -2{\cal R}_{AB}n^{A}n^{B}
    +K^{2}-K_{\mu\nu}K^{\mu\nu}~,
\label{gauss}
\end{equation}
and the Codazzi relation
\begin{equation}
    (Kg^{\mu\nu}-K^{\mu\nu})_{;\nu}=n^{A}{\cal R}_{AB}y^{B}_{\,,\nu}g^{\mu\nu}~.
\label{coddazi}
\end{equation}
The variation of the unit normal vector (\ref{normal}) is
$\delta n_{a}=\frac{\delta \gamma}{\gamma}n_{a}+\gamma(\delta f)_{,a}$.
But the brane is in rest in the $z$ system, therefore $\delta f=0$,
and the variation of the
normal unit vector (\ref{normal}) is only in its magnitude
not in the direction
\begin{equation}
    \delta n_{a} = -\frac{1}{2}n_{a}(n_{b}n_{c}\delta G^{bc})~.
    \label{varn}
\end{equation}
Looking first only at the gravitational action
on one side of the brane
\begin{equation}
    16\pi{\bf G}I_{g} = \int_{V}d^{N+1}z\,\sqrt{-G}{\cal R}
        -2\int_{B}d^{N}x\,\sqrt{-g}K \label{gravaction} ,
\end{equation}
the relevant variations are
\begin{itemize}
\item The variation of the volume integral
\begin{equation}
    \delta(\sqrt{-G}{\cal R})=\sqrt{-G}\left[({\cal R}_{ab}
    -\frac{1}{2}{\cal R}G_{ab})\delta G^{ab}
    +(G^{ab}\delta\Gamma^{c}_{ab}-G^{ac}\delta\Gamma^{b}_{ab})_{;c} \right]
    \label{varR}
\end{equation}
\item The variation of the brane integral involves the variation of
$\sqrt{-g}$
\begin{equation}
    \delta\sqrt{-g}=\frac{1}{2}\sqrt{-g}g^{\mu\nu}\delta g_{\mu\nu}=
    -\frac{1}{2}\sqrt{-g}h_{ab}\delta G^{ab} \label{varg},
\end{equation}
and the variation of the extrinsic curvature (\ref{excurv})
\begin{equation}
    G^{ab}\delta K_{ab}= -\frac{1}{2}Kn_{a}n_{b}\delta G^{ab}
    + n_{c}h^{ab}\delta\Gamma^{c}_{ab} ~.
    \label{varK}
\end{equation}
where $h_{ab}=G_{ab}-n_{a}n_{b}$ is the first fundamental form (\ref{fff})
and we have made use of (\ref{varn},\ref{fff},\ref{excurv}).
\end{itemize}
Using Gauss law for an integral over a total divergence
\begin{equation}
    \int_{V}d^{N+1}z\,\sqrt{-G}V^{a}_{\,;a}
=\int_{B}d^{N}x\,\sqrt{-g}V^{a}n_{a}
    \label{divergence}
\end{equation}
the total variation of the gravitational action (\ref{gravaction})
is simply
\begin{eqnarray}
    16\pi{\bf G}\delta I_{g} &=& \int_{V}d^{N+1}z\,\sqrt{-G}
    \left[({\cal R}_{ab}-\frac{1}{2}{\cal R}G_{ab})\delta G^{ab}\right]
    \nonumber \\
    &+& \int_{B}d^{N}x\,\sqrt{-g}\left[-2(K_{ab}-\frac{1}{2}KG_{ab})
    \delta G^{ab}
    + (-2n_{c}h^{ab}+n_{c}G^{ab}-n^{a}\delta^{b}_{c})\delta\Gamma^{c}_{ab}
    \right] ~.
    \label{vargrav}
\end{eqnarray}
Using $G_{ab}=h_{ab}+n_{a}n_{b}$ (\ref{fff}),
expressing $\delta\Gamma^{c}_{ab}$ in terms of $(\delta G^{ab})_{;c}$,
and using the integral relation
\begin{equation}
    \int_{B}d^{N}x\,\sqrt{-g}V^{a}_{\,;b}h^{b}_{a}
    =-\int_{B}d^{N}x\,\sqrt{-g}V^{a}n_{a}K ,
    \label{integral2}
\end{equation}
the term proportional to $\delta\Gamma^{c}_{ab}$
in the brane integral in Eq.(\ref{vargrav}) is turned into
$(K_{ab}-Kn_{a}n_{b})\delta G^{ab} $. The total variation of
the gravitational action on one side (\ref{gravaction}) is given by
\begin{eqnarray}
    16\pi{\bf G}\delta I_{g} &=& \int_{V}d^{N+1}z\,\sqrt{-G}
    ({\cal R}_{ab}-\frac{1}{2}{\cal R}G_{ab})\delta G^{ab}
    \nonumber \\
    &+& \int_{B}d^{N}x\,\sqrt{-g}(Kh_{ab}-K_{ab})
    \delta G^{ab}~.
    \label{vargrav2}
\end{eqnarray}
The total variation of the action(\ref{genaction}) has a contribution
from the gravitational action integral on each side of the brane, and
matter contributions from the various Lagrangians.
\begin{eqnarray}
    16\pi{\bf G}\delta I &=& \int_{V_{1}}d^{N+1}z\,\sqrt{-G}
    \left[({\cal R}_{ab}-\frac{1}{2}{\cal R}G_{ab} -8\pi{\bf G}
    {\cal T}_{ab})\delta G^{ab}\right]
    \nonumber \\
    &+& \int_{V_{2}}d^{N+1}z\,\sqrt{-G}
    \left[({\cal R}_{ab}-\frac{1}{2}{\cal R}G_{ab} -8\pi{\bf G}
    {\cal T}_{ab})\delta G^{ab}\right]  \label{totalvar} \\
    &+& \int_{B}d^{N}x\,\sqrt{-g}\left[\left((Kh_{ab}-K_{ab})
    \delta G^{ab}\right)^{(1)} +\left((Kh_{ab}-K_{ab})
    \delta G^{ab}\right)^{(2)} -8\pi{\bf G}T_{\mu\nu}
    \delta g^{\mu\nu}
    \right] ~.\nonumber
\end{eqnarray}
Here $\displaystyle{{\cal T}^{(1,2)}_{ab}=-\frac{2}{\sqrt{-G}}
\frac{\delta(\sqrt{-G}{\cal L}^{(1,2)})}{\delta G^{ab}}}$
is the bulk energy- momentum tensor,
and $\displaystyle{T_{\mu\nu}=-\frac{2}{\sqrt{-g}}
\frac{\delta(\sqrt{-g}L_{B})}{\delta g^{\mu\nu}}}$ is the
energy-momentum tensor of the brane
\footnote{The brane action may include the intrinsic curvature
of the brane, this is the Einstein-Hilbert action. In that case,
the energy momentum tensor of the brane includes the
Einstein tensor
$$T_{\mu\nu}\rightarrow T_{\mu\nu}-\frac{1}{8\pi G_{N}}(R_{\mu\nu}
-\frac{1}{2}g_{\mu\nu}R)$$.}.

\subsection{The Equations of Motion}

The next step is crucial in Dirac's method. One has to express
the variation of the metric in the $z$ system in terms of the variations
in the $y$ system. Using Eq.(\ref{zmetric})
\begin{equation}
    \delta G_{ab}=\delta G_{AB}y^{A}_{\,,a}y^{B}_{\,,b}
    +2\left(G_{AB}y^{A}_{\,,a}\delta y^{B}\right)_{;b}~.
\end{equation}
The contraction of a general tensor $F_{ab}$ with
the variation of the metric $\delta G^{ab}$ is
\begin{equation}
    F_{ab}\delta G^{ab} = -F^{ab}\delta G_{ab} =
     F_{AB}\delta G^{AB} +2(F^{AB})_{;B}G_{AD}\delta y^{D}
    -2\left(F^{ab}G_{AB}y^{A}_{\,,a}\delta y^{B}\right)_{\,;b} ~.
    \label{noether}
\end{equation}
After substituting (\ref{noether}) in (\ref{totalvar}) and using
(\ref{divergence}) to turn the total divergence into a surface integral,
we can transform back to the $y$ system and we are left with
\begin{eqnarray}
    16\pi{\bf G}\delta I &=& \int_{V_{1}}d^{N+1}y\,\sqrt{-G}
    \left[{\cal E}_{AB}\delta G^{AB}
    +2({\cal E}_{A}^{B})_{;B}\delta y^{A}\right]
    \nonumber \\
    &+& \int_{V_{2}}d^{N+1}y\,\sqrt{-G}
\left[{\cal E}_{AB}\delta G^{AB}
    +2({\cal E}_{A}^{B})_{;B}\delta y^{A}\right]
\nonumber \\
    &+& \int_{B}d^{N}x\,\sqrt{-g}\left[\left((Kh_{ab}-K_{ab})
    \delta G^{ab}\right)^{(1)} +\left((Kh_{ab}-K_{ab})
    \delta G^{ab}\right)^{(2)} -8\pi{\bf G}T_{\mu\nu}
    \delta g^{\mu\nu}\right. \nonumber   \\
    &\;\;\;- & \left. 2\left(n^{A}{\cal E}_{AB}\delta y^{B}\right)^{(1)}
    -2\left(n^{A}{\cal E}_{AB}\delta y^{B}\right)^{(2)}
    \right] ~.\label{diracvar}
\end{eqnarray}
Where ${\cal E}_{AB}={\cal R}_{AB}-\frac{1}{2}{\cal R}G_{AB} -8\pi{\bf G}
    {\cal T}_{AB}$ is Einstein's factor in the surrounding manifold.

\begin{itemize}
\item The equations of motion in the surrounding manifold are clear.
    The arbitrariness of $\delta G^{AB}$ leads to Einstein's equation
\begin{equation}
    {\cal R}_{AB}-\frac{1}{2}{\cal R}G_{AB} -8\pi{\bf G}
    {\cal T}_{AB} =0 ~. \label{ein}
\end{equation}
While the arbitrariness of $\delta y^{A}$ combined with the Bianchi identity
will lead to energy-momentum
conservation, which is the conserved Noether current associated with
general coordinate transformation
\begin{equation}
    ({\cal T}^{AB})_{;B} =0 ~. \label{emconserv}
\end{equation}
\item
The crucial question is what are the exact relations between the variations
on the brane. The most important thing is that the brane metric $g_{\mu\nu}$
will be well defined.
Therefore the variations of $G_{AB}(y)$ and $y^{A}(z)$ are not
independent, but are constrained in such a way that
\begin{equation}
    \delta G^{ab(1)}=\delta G^{ab(2)}
    =z^{a}_{\,,\mu}z^{b}_{\,,\nu}\delta g^{\mu\nu}
    \label{deltag}
\end{equation}
This will lead to the Israel condition \cite{israel}
\begin{equation}
     \fbox{$\displaystyle{
    (Kg_{\mu\nu}-K_{\mu\nu})^{(1)} +(Kg_{\mu\nu}-K_{\mu\nu})^{(2)}
    =8\pi{\bf G}T_{\mu\nu}
     }$}
    \label{WI}
\end{equation}
Notice that in our notation the outward pointing normal vector
has a general sign change from one side to the other, and the above
expression is actually the difference in extrinsic curvature.
\item
The variation in the location of the brane can be projected tangent
and normal to the brane. The tangent variation should be continuous
over the brane
\begin{equation}
    (\delta y^{A}G_{AB}y^{B}_{\,,\mu})^{(1)}
    =(\delta y^{A}G_{AB}y^{B}_{\,,\mu})^{(2)} ~.
    \label{tanvar}
\end{equation}
The equation resulting from (\ref{diracvar}) with the arbitrariness of
(\ref{tanvar}) is
\begin{equation}
    \left(n^{A}({\cal R}_{AB}-\frac{1}{2}{\cal R}G_{AB} -8\pi{\bf G}
    {\cal T}_{AB})y^{B}_{\,,\mu} \right)^{(1)}
    +\left(n^{A}({\cal R}_{AB}-\frac{1}{2}{\cal R}G_{AB} -8\pi{\bf G}
    {\cal T}_{AB})y^{B}_{\,,\mu} \right)^{(2)} =0~.
    \label{tan1}
\end{equation}
To understand that equation, one should notice that $n_{A}y^{A}_{\,,\mu}=0$,
and use Codazzi's equation (\ref{coddazi}) and the Israel condition (\ref{WI})
to get the energy momentum conservation on the brane
\begin{equation}
     \fbox{$\displaystyle{
    (T_{\mu}^{\nu})_{;\nu}
    =(n^{A}{\cal T}_{AB}y^{B}_{\,,\mu})^{(1)}
    +(n^{A}{\cal T}_{AB}y^{B}_{\,,\mu})^{(2)}~.
     }$}
    \label{Bemcnsrv}
\end{equation}
The total change in energy-momentum confined to the brane is due to
flow of energy momentum in and out of the brane.
\item
The normal variations in the location of the brane are opposite in sign
in our notations $(n_{A}\delta y^{A})^{(1)}=-(n_{A}\delta y^{A})^{(2)}$.
The arbitrariness of the normal variation in (\ref{diracvar}) will lead to
\begin{equation}
    \left(n^{A}({\cal R}_{AB}-\frac{1}{2}{\cal R}G_{AB} -8\pi{\bf G}
    {\cal T}_{AB})n^{B} \right)^{(1)}
    -\left(n^{A}({\cal R}_{AB}-\frac{1}{2}{\cal R}G_{AB} -8\pi{\bf G}
    {\cal T}_{AB})n^{B} \right)^{(2)} =0~.
    \label{normal1}
\end{equation}
Using Gauss relation (\ref{gauss}), and substituting Israel condition
(\ref{WI}) multiplied by \protect{$(K_{\mu\nu}^{(1)}-K_{\mu\nu}^{(2)})$}
to get the non homogenous geodetic brane equation
\begin{equation}
     \fbox{$\displaystyle{
    \frac{1}{2}T^{\mu\nu}(K_{\mu\nu}^{(1)}-K_{\mu\nu}^{(2)})
    ={\cal T}_{nn}^{(1)}-{\cal T}_{nn}^{(2)}~.
     }$}
    \label{GB}
\end{equation}
This is the analog of Newton's second law. The average value of the
extrinsic curvature is the acceleration and the energy momentum of the
brane is the analog of mass. The right hand side is the net force
(pressure) acting on the brane
\footnote{Let us emphasize that Eq.(\ref{GB}) is a consequence
of Eq.(\ref{ein},\ref{WI}) if the bulk gravitational coupling
is strong, but, Eq.(\ref{GB}) remains valid even in the case
where the bulk gravitational coupling vanishes.}.
\end{itemize}

\section{Summary}
The equations of the brane were derived using the variational
principle. The action is that of a brane embedded in co-dimension one.
These equations are not new and were derived in the past using other methods
\cite{israel,carter}. Dirac's method is essential in the derivation of
Eqs.(\ref{Bemcnsrv}, \ref{GB}). While Israel condition can be read off
Eq.(\ref{totalvar}), energy momentum conservation on the brane (\ref{Bemcnsrv}) and the
geodetic equation of the brane (\ref{GB}) emerge only when using Dirac's method.

In addition, Israel junction condition is a remnant of Einstein equations
in the bulk. It is relevant only when the bulk gravitational coupling is strong.
On the other hand, energy momentum conservation on the brane (\ref{Bemcnsrv}) and the
geodetic equation of the brane (\ref{GB}) are the equations of the brane. These
are valid even if the bulk gravitational coupling vanishes \cite{RT}.

\end{document}